\documentclass[prd,longbibliography
,groupedaddress, nofootinbib,notitlepage
]{revtex4-1}

\usepackage{amsmath,amssymb}
\usepackage{natbib}
\usepackage{graphicx}

\newcommand{\be}{\begin{equation}}
\newcommand{\ee}{\end{equation}}
\newcommand{\bse}{\begin{subequations}}
\newcommand{\ese}{\end{subequations}}
\newcommand{\ba}{\begin{eqnarray}}
\newcommand{\ea}{\end{eqnarray}}
\newcommand{\bea}{\begin{eqnarray}}
\newcommand{\eea}{\end{eqnarray}}

\newcommand{\ts}{t_\star}
\newcommand{\rhos}{\rho_\star}

\newcommand{\mn}{{\mu\nu}}

\usepackage{color}
\usepackage[normalem]{ulem}  

\begin{document}


\title{Hydrodynamic attractor and novel fixed points in superfluid Bjorken flow}


\author{Toshali Mitra}
\email[]{toshalim@imsc.res.in}
\affiliation{The Institute of Mathematical Sciences, Chennai 600113, India}

\author{Ayan Mukhopadhyay}
\email[]{ayan@physics.iitm.ac.in}
\affiliation{Department of Physics, Indian Institute of Technology Madras,\\ Chennai 600036, India}

\author{Alexander Soloviev}
\email[]{alexander.soloviev@stonybrook.edu}
\affiliation{Center for Nuclear Theory, Department of Physics and Astronomy,\\ Stony Brook University, Stony Brook, New York 11794, USA}

\date{\today}

\begin{abstract}
Extending the quantum effective approach of Son and Nicolis and incorporating dissipation, we develop a MIS formalism for describing a superfluid out of equilibrium by including the Goldstone boson and the condensate together with the hydrodynamic modes as the effective degrees of freedom. We find that the evolution of the superfluid undergoing Bjorken flow is governed by the conventional hydrodynamic attractor with unbroken symmetry	 and an even number of novel non-dissipative fixed points with broken symmetry. If the initial temperature is super-critical, then the condensate becomes exponentially small very rapidly and the system is trapped by the hydrodynamic attractor for a long intermediate time before it reheats rapidly and switches to one of the symmetry-breaking fixed points eventually. \textcolor{black}{Finally, we show that the fixed points are unstable against inhomogeneous perturbations that should lead to spinodal decomposition.} We conclude that these features should be generic beyond the MIS formalism.
\end{abstract}

\maketitle

\section{Introduction}
Phase transitions in out-of-equilibrium systems are of interest for the physics of heavy-ion collisions \cite{Teaney:2009qa,Berges:2020fwq}, cold atoms \cite{Adams:2012th} and the early Universe \cite{Mazumdar:2018dfl}.  Given the success of hydrodynamic theories for understanding out-of-equilibrium matter with both weak and strong self-interactions, there is sufficient motivation to construct frameworks that can describe superfluids by incorporating Goldstone mode(s) and the order parameter(s) (condensate(s)) associated with spontaneous breaking of global internal symmetries in hydrodynamic theories.  Recently the demonstration of the ubiquitous presence of \textit{hydrodynamic attractors} in a wide variety of phenomenological frameworks (eg. kinetic theory, holography, etc) \textcolor{black}{
has given us insights into how a system \textit{hydrodynamizes} far away from equilibrium such that first order hydrodynamics can describe the evolution of the energy-momentum tensor and conserved currents even in presence large pressure gradients \cite{Heller:2015dha,Romatschke:2017vte,Romatschke:2017ejr,Florkowski:2017olj}.}  It is therefore pertinent to ask if one can construct 
similar frameworks that incorporate Goldstone boson(s) and order parameter(s) along with hydrodynamic modes for superfluid states far from equilibrium.

A general framework for superfluids at finite density has its roots in the pioneering work of Son \cite{Son:2002zn} (and generalized to finite temperature by Nicolis \cite{Nicolis:2011cs}), which utilized the quantum effective action for describing the low energy dynamics of the Goldstone bosons and superfluid vortices
. An 
attractive feature of Son's approach is that the only microscopic input needed 
is simply the equation of state of the system at finite chemical potential 
 and temperature (see \cite{Glorioso:2018wxw} for a survey of related approaches and a broader discussion). 
   Remarkably, the thermodynamic description emerges naturally from the energy-momentum tensor and conserved currents obtained from the quantum effective action and is not imposed by hand.  The original context in which Son developed this approach was to describe baryonic matter at finite baryon density with spontaneous breaking of $U(1)$ baryon symmetry relevant for physics of neutron stars.  Recently, generalizations of the quantum effective action have been studied incorporating both the pions and the hydrodynamic modes at finite chemical potential and temperature \cite{Grossi:2020ezz}. Further extensions to the case of phases with spontaneous spacetime symmetry breaking such as supersolids, etc \cite{Watanabe:2013iia,Nicolis:2013lma,Alberte:2015isw,Nicolis:2015sra} have been investigated as well but this is beyond the scope of this work.

In this work, we include the order parameter and add dissipation by merging the Son-Nicolis framework with the M{\"u}ller-Israel-Stewart (MIS) formalism \cite{Muller:1967zza,Israel:1979wp} enabling \textcolor{black}{a causal description. We use this to} study symmetry breaking in out-of-equilibrium situations in the context of Bjorken flow. The study of evolution of many-body systems under Bjorken flow is particularly of interest not only for heavy-ion physics \cite{Romatschke:2017ejr}, but also for cold atomic systems in anisotropic laser traps \cite{Mitra:2018xob,Glidden:2020qmu}. We restrict ourselves here to the case of a $U(1)$ symmetry breaking but our work can be easily generalized to non-Abelian cases.

At early time, the size of the expanding system undergoing Bjorken flow is sufficiently small so that fluctuations are significant. Therefore we should study the evolution of the system \textcolor{black}{with generic initial conditions for the energy-momentum tensor, conserved currents and the condensate.} Remarkably, we find that as long as we start above the critical temperature, $T_c$, the condensate decays exponentially while the phase evolves over a similar timescale to satisfy the Josephson condition. Then the full system gets trapped very close to a conventional \textit{hydrodynamic attractor} 
over a very long period of time during which the system approaches a perfect fluid expansion with unbroken symmetries. 

However, the long time physics is somewhat surprising. It turns out that the superfluid system has an even number of symmetry breaking non-dissipative fixed points in which the full system undergoes expansion at a \textit{constant} temperature (determined by the equation of state and the potential) and with a \textit{constant} value of the condensate. These fixed points are possible because the condensate lowers the energy with respect to the vacuum, allowing the expanding system to maintain a self-consistent constant temperature (with a non-trivial velocity profile). We find that with inital 
$T>T_c$ and otherwise generic initial conditions, the superfluid system switches rapidly to one of these symmetry breaking fixed points after spending a  time in very close proximity to the conventional \textit{hydrodynamic attractor}. The basin of attraction of these fixed points has complicated interlacing and possibly fractal boundaries. If $T< T_c$ initially, then the superfluid system generically evolves to one of the fixed points without getting trapped near the hydrodynamic attractor (unless the initial condition is close to the boundary between the basins of attraction of the fixed points).

Our work demonstrates that the evolution of superfluid matter out of equilibrium is governed by the conventional hydrodynamic attractor with unbroken symmetry (which is rather like a saddle curve in the extended phase space) and non-trivial symmetry breaking fixed points which are determined by the  potential of the condensate and the equation of state. The hydrodynamic attractor governs physics at intermediate time scales if the initial temperatures are above $T_c$, while the fixed points govern the physics at much longer time scales with a (typically) rapid crossover. 

We observe that the symmetry breaking fixed points are independent of the relaxation mechanism, and are determined only by the equation of state and the potential of the condensate. Furthermore, the hydrodynamic attractor is a feature in any phenomenological framework incorporating relaxation as mentioned above. Therefore, the dynamical features of superfluid flow, especially with respect to the role of the hydrodynamic attractor and the fixed point should be universal. 

However, if there is competition between various types of symmetry breaking as in neutron stars and some strongly correlated systems, we will need to incorporate a more general framework discussed in \cite{Alford:2012vn,Alford:2013ota,Haber:2015exa} that can couple multiple (super)fluids. This is beyond the scope of our present work. \textcolor{black}{We also postpone a discussion regarding the relevance of our results for the quark-gluon plasma until the concluding section.}  

\textcolor{black}{We also analyze the linearized perturbations around thermal equilibrium and show that the fixed points are unstable against inhomogeneous perturbations that should lead to spinodal decomposition. This can be studied using our MIS framework, which we postpone this to a future work.}

The organization of the paper is as follows. In Section \ref{setup}, we discuss the construction of the MIS theory of superfluids. In Section \ref{Bjorken}, we derive the equations for Bjorken flow and then present our results in Section \ref{results}. \textcolor{black}{We further analyze the system by considering linearized perturbations in Section \ref{lin-pert}}. Finally we conclude with a discussion on the limitations of our results and necessary future work in Section \ref{discussion}.

\section{The MIS formulation of superfluid effective theory}\label{setup}

\subsection{\textcolor{black}{Effective action}}

The MIS formalism 
\cite{Muller:1967zza,Israel:1979wp} 
enables a causal description of a relativistic hydrodynamic system. In order to generalize this approach to a relativistic superfluid, we utilize aspects of the effective theory of superfluids due to Son and Nicolis \cite{Son:2002zn,Nicolis:2011cs} 
at finite temperature. Furthermore, we include the order parameter as a dynamical variable to incorporate the possibility of out-of-equilibrium phase transition. 

For simplicity, consider a complex field $\Sigma \equiv \rho e^{i\psi}$ charged under $U(1)$ global symmetry which is broken spontaneously below a critical temperature $T_c$ due to the potential $V(\rho, \dots)$ where $\dots$ denote other variables of the effective theory.  Crucially, the velocity $u^\mu$ and temperature $T$ of the local thermal frame appear together as a new fundamental variable $\beta^{\mu}$.  Note that in this approach (local) thermodynamics emerge from the effective action and is not imposed separately. We first succinctly show how this works.

Due to Lorentz invariance, the effective action in a background metric $g_{\mu\nu}$ and gauge field $A_\mu$ (\textcolor{black}{introduced to allow us to compute the energy-momentum tensor and conserved current efficiently}) can depend on $\beta^\mu$ only via the \textit{temperarture} $T$ and \textit{chemical potential} $\mu$ defined by the relations $ T := (-\beta^\mu g_{\mu\nu} \beta^\nu)^{-1/2}$ and  $\mu/T := A_\mu \beta^\mu$, respectively, at leading order in derivatives. Similarly, due to $U(1)$ gauge and Lorentz invariance, the dependence on $\Sigma$ can be via the scalars $X := (D_\mu \Sigma)^\dagger (D^\mu \Sigma)$ and $Y:= ((\beta \cdot D) \Sigma)^\dagger ((\beta \cdot D)  \Sigma)$, where 
\begin{align}
D_\mu \Sigma = (\nabla_\mu + i A_\mu) \Sigma = (\partial_\mu \rho +  i \rho D_\mu \psi) e^{i\psi},
\end{align}
with $D_\mu \psi := \nabla_\mu \psi + A_\mu$ and 
\textcolor{black}{$\nabla_\mu$ is the covariant derivative constructed from the background metric}. It will turn out (from absence of entropy production) that $(u\cdot D) \psi = 0$, i.e. $(u\cdot \nabla) \psi  = \mu$ (a.k.a. the Josephson condition) must be satisfied in equilibrium while $\rho = \rho_*$, a constant, so that $Y$ must vanish. On the other hand, we will find that $X$ can assume an arbitrary non-trivial profile in equilibrium due to non-vanishing of derivatives of $\psi$ in directions orthogonal to the \textcolor{black}{four-velocity $u^\mu\equiv T\beta^\mu $}.  Since $Y$ must vanish at equilibrium, we will drop it and retain $X$, $T$ and $\mu$ as fundamental variables. Additionally we incorporate the gauge-invariant variable $\rho$ to implement spontaneous symmetry breaking. Since derivatives of $\rho$ (and $\psi$) appear in $X$, we must think of $X$ and $\rho$ as independent variables.\footnote{\textcolor{black}{Note that as in a classical scalar field theory, $\rho$ and its conjugate field momentum should also be thought of as independent variables since we need to set their initial conditions independently to specify an evolution. This implies that $X$ and $\rho$ should be thought of as independent variables.}} 

We therefore consider an effective action of the form: 
\begin{equation}\label{Eq:EffectiveAction}
S
=\int d^4 x \sqrt{-g}\left[ F(X,T, \mu) - V(\rho, T, \mu)\right]\,,
\end{equation}
where $F$ is the kinetic piece of the action and $V$ is the symmetry breaking potential. Simple instances of $F$ and $V$ are
\begin{align}\label{Eq:FV}
F = -\frac{1}{2}X + p(T,\mu), \quad V(\rho,  T, \mu) = \frac{\sigma (T-T_c)}{2} \rho^2 + \frac{\lambda}{4} \rho^4
\end{align}
in which $\sigma,\lambda >0$ and are both independent of $T$ and $\mu$.  In Son's approach where only the dynamics of the Goldstone mode are considered \cite{Son:2002zn}, the condensate is set to its equilibrium value. The dependence of the generalized pressure $F$ on the chemical potential $\mu$ then also determines its $X$ dependence. In our case, the condensate is an independent variable and we truncate our theory to two-derivatives. To relate to Son's approach, we can simply expand the generalized pressure $F$ in derivatives of the condensate and normalize the latter to get a standard two-derivative kinetic term. This leads to our simple ansatz \eqref{Eq:FV}. In what follows for the rest of this section, we consider a general $F(X, T,\mu)$.


\subsection{\textcolor{black}{Ideal hydrodynamics and thermodynamics}}

The \textit{ideal} \textcolor{black}{superfluid} energy-momentum tensor and conserved $U(1)$ current can be readily obtain by varying the action \eqref{Eq:EffectiveAction} w.r.t. $g_{\mu\nu}$ and $A_\mu$ respectively:
\begin{align}
T^{\mu\nu}_{\rm ideal} \equiv\frac{2}{\sqrt{-g}}\frac{\delta S}{\delta g_{\mu\nu}}&= - 2 \frac{\partial F}{\partial X}\left(\nabla^\mu \rho \nabla^\nu \rho + \rho^2 D^\mu \psi D^\nu \psi\right) \nonumber\\\label{Eq:idealemt}
&+(F- V) \Delta^{\mu\nu} +  u^\mu u^\nu\left(T \frac{\partial}{\partial T}  +\mu \frac{\partial}{\partial \mu} -1\right) (F-V),\\
\label{Eq:idealj}
j^\mu_{\rm ideal} \equiv\frac{1}{\sqrt{-g}}\frac{\delta S}{\delta A_{\mu}} &= \frac{\partial (F- V)}{\partial \mu}  u^\mu +  \rho^2  D^\mu \psi,
\end{align}
respectively with $\Delta^{\mu\nu} = g^{\mu\nu} + u^\mu u^\nu$. Each of these can be rewritten as sum of a normal and a coherent superfluid component. Explicitly,
\begin{eqnarray}\label{Eq:idealemt2}
T^{\mu\nu}_{\rm ideal} = T^{\mu\nu}_n- 2 \frac{\partial F}{\partial X}\left( \nabla^\mu \rho \nabla^\nu \rho + \rho^2 D^\mu \psi D^\nu \psi\right),\quad T^{\mu\nu}_n :=\mathcal{E} u^\mu u^\nu + P \Delta^{\mu\nu},\quad   j^\mu_{\rm ideal} = j^\mu_n +  j^\mu_\psi, \quad j^\mu_n := \mathcal{N} u^\mu.
\end{eqnarray}
where $j^\mu_\psi := \rho^2 D^\mu \psi$ and
\begin{eqnarray}\label{Eq:ePn}
P := F - V, \quad  \mathcal{E}  := \left(T \frac{\partial}{\partial T}  +\mu \frac{\partial}{\partial \mu} -1\right) (F-V), \quad \mathcal{N}  := \frac{\partial (F-V)}{\partial \mu} =\frac{\partial P}{\partial \mu} .
\end{eqnarray}
With the entropy density $\mathcal{S} $ defined as
\begin{equation}\label{Eq:s}
\mathcal{S}  := \frac{\partial P}{\partial T}{\rm  \Big \vert_{\mu \text{ fixed}}},
\end{equation}
the relations \eqref{Eq:ePn} imply the standard thermodynamic identity 
\begin{equation}
P= -\mathcal{E}  + T\mathcal{S}  + \mu \mathcal{N} . 
\end{equation}
Note that $P$ is distinct from $p$ which appears in \eqref{Eq:FV} 
\textcolor{black}{as it is the generalized pressure which governs the normal fluid component. The coherent part in \eqref{Eq:idealemt} which is proportional to $D^\mu \psi D^\nu\psi$ can be reinterpreted as a second fluid with a velocity field proportional to $D^\mu\psi$ and with an independent free energy. We refer the reader to \cite{Alford:2012vn,Alford:2013ota,Schmitt:2014eka} for more details. Note in our case we will also need a third fluid component with velocity field proportional to $\nabla^\mu \rho$ as evident from \eqref{Eq:idealemt}. However, such reinterpretations will not be necessary in what follows.}

Varying the effective action \eqref{Eq:EffectiveAction} w.r.t. $\Sigma$, we obtain the equations of motion for $\rho$ and $\psi$ which are
\begin{equation}\label{Eq:EomSigma1}
2\nabla_\mu \left(\frac{\partial F}{\partial X}\nabla^\mu \rho\right)  +\frac{\partial V}{\partial \rho} -2 \rho \frac{\partial F}{\partial X} D_\mu\psi D^\mu\psi = 0, \quad \text{and} \quad \nabla_\mu j^\mu_\psi = 0.
\end{equation}
The above equations of motion, in conjunction with $\nabla_\mu T^{\mu\nu}_{\rm ideal} = 0$ and $\nabla_\mu j^{\mu}_{\rm ideal} = 0$, imply the Euler equations
\begin{eqnarray}
\nabla_\mu (\mathcal{S} u^\mu) = 0, \quad \nabla_\mu (\mathcal{N}  u^\mu) = 0, \quad (T\mathcal{S}  + \mu \mathcal{N} )(u\cdot \nabla) u^\mu = - \left( \mathcal{S}  \nabla_\perp^\mu  T + \mathcal{N} {\nabla_\perp}^\mu \mu\right)
\end{eqnarray}
where $\nabla_\perp^\mu \equiv \Delta^{\mu\nu}\nabla_\nu$.

\subsection{\textcolor{black}{Adding dissipation}}
In order to further extend the effective theory, we need to add \textcolor{black}{dissipation. This is achieved by adding additional terms to the energy-momentum tensor and the conserved current, which we denote as $\pi^{\mu\nu}$ and $q^\mu$ respectively, and which can be expanded in the form of derivatives of the Goldstone, velocity and temperature fields. The latter are called \textit{constitutive relations} which should be determined/constrained by enforcing the existence of an entropy current with non-negative divergence. We therefore begin by writing}  
\begin{equation}\label{Eq:FullTj}
T^{\mu\nu}_n =\mathcal{E}u^\mu u^\nu +P g^{\mu\nu}+ \pi^{\mu\nu}, \quad  T^{\mu\nu} =T^{\mu\nu}_n - 2 \frac{\partial F}{\partial X}\left( \nabla^\mu \rho \nabla^\nu \rho + \rho^2 D^\mu \psi D^\nu \psi\right), \quad j^{\mu}_n = \mathcal{N} u^\mu+ q^{\mu}, \quad j^\mu = j^{\mu}_n + j^\mu_\psi
\end{equation}
with $\mathcal{E}$, $P$ and $\mathcal{N}$ as defined in \eqref{Eq:ePn}. We choose a generalization of the Landau frame to define $u^\mu$ in the non-ideal case, so that $\pi^{\mu\nu} u_\nu = 0$. 
Furthermore, we modify \eqref{Eq:EomSigma1} by adding dissipative sources $\theta_1$ and $\theta_2$ to the equations of motion of $\rho$ and $\psi$ respectively, so that
\begin{equation}\label{Eq:EomSigma2}
2\nabla_\mu \left(\frac{\partial F}{\partial X}\nabla^\mu \rho\right)  +\frac{\partial V}{\partial \rho} -2 \rho \frac{\partial F}{\partial X} D_\mu\psi D^\mu\psi = \theta_1, \quad \nabla_\mu j^\mu_\psi = \theta_2.
\end{equation}
The above in conjunction with $\nabla_\mu T^{\mu\nu}= 0$ and $\nabla_\mu j^{\mu} = 0$ imply
\begin{eqnarray}
T\nabla_\mu\left(\mathcal{S}u ^\mu  \right) &=& -\theta_1 (u\cdot\nabla) \rho - (\mu + (u\cdot D)\psi)\theta_2 - \pi^{\mu\nu}(\nabla_\mu u_\nu )+\mu (\nabla\cdot q), \\ 
(\mathcal{S} T +\mathcal{N} \mu )(u\cdot\nabla) u^\mu &=& -\mathcal{S} {\nabla_\perp}^\mu T - \mathcal{N} {\nabla_\perp}^\mu \mu  -{\nabla_\perp}_\nu \pi^{\mu\nu} + \theta_1 {\nabla_\perp}^\mu \rho + \theta_2 {D_\perp}^\mu \psi
\end{eqnarray}
where ${D_\perp}^\mu = \Delta^{\mu\nu}D_\nu$. As a consequence, we obtain a candidate entropy current $j^\mu_s = \mathcal{S} u ^\mu - (\mu/T)  (q^\mu+ j^\mu_\psi)$ whose divergence turns out to be
\begin{equation}\label{Eq:EntropyCurrent}
\nabla_\mu j^\mu_s = - \frac{1}{T}(\nabla_\mu u_\nu) \pi^{\mu\nu} -((q + j_\psi)\cdot \nabla)  \frac{\mu}{T}  - \frac{1}{T}\theta_1 (u\cdot\nabla) \rho -\frac{1}{T}\theta_2 (u\cdot D) \psi.
\end{equation}
The appropriate constitutive relations that lead to the positive definite divergence of the entropy current is then
\begin{eqnarray}\label{Eq:Constitutive}
\pi^{\mu\nu} = -2 \eta \sigma^{\mu\nu} - \zeta \Delta^{\mu\nu} (\nabla\cdot u), \quad
q^\mu  + j^\mu_\psi =  - \kappa \nabla^\mu \left( \frac{\mu}{T}\right), \quad
\theta_1 = -\kappa_1 (u\cdot\nabla) \rho, \quad \theta_2 = - \kappa_2 (u\cdot D) \psi
\end{eqnarray}
with $\kappa$, $\eta$, $\zeta$, $\kappa_1$ and $\kappa_2$ positive definite functions of $T$ and $\mu$ (and in principle of $\rho$ too). Also  $$\sigma^\mn :=\frac{1}{2} \Delta^{\mu\alpha}\Delta^{\nu\beta}(\nabla_\alpha u_\beta+\nabla_\beta u_\alpha)-\frac{1}{3}\Delta^\mn\nabla_\alpha u^\alpha$$ is the shear-stress tensor. We note that the constitutive relation for $q^\mu$ implies that the conserved $U(1)$ current is
\begin{equation}
j^\mu = n u^\mu  -\kappa \nabla^\mu \left( \frac{\mu}{T}\right).
\end{equation}
We readily validate our earlier claim from \eqref{Eq:EntropyCurrent} and \eqref{Eq:Constitutive} that in equilibrium where entropy production is absent, we should satisfy the Josephson condition $(u\cdot D) \psi = 0$, i.~e. $(u\cdot \nabla) \psi = \mu$. Furthermore, \eqref{Eq:EomSigma2} implies that at equilibrium $\rho$ should be in the (thermal) vacuum so that $\partial V/\partial \rho = 0$ in absence of spatial gradients of $\psi$. Note when ${\nabla_\perp}_\mu \psi {\nabla_\perp}^\mu \psi \neq 0$, e.~g. a constant, we can have $\rho$ taking values away from the minima of $V$ in equilibrium at any $T$ and $\mu$.

Finally in order to develop a MIS-type formulation, we simply replace the constitutive relations for $\pi^{\mu\nu}$ and $j^\mu$ in \eqref{Eq:Constitutive} by the dynamical equations
\begin{eqnarray}\label{Eq:MIS}
(u\cdot\nabla)(q^\mu+ j^\mu_\psi)+ \frac{1}{\tau_q}(q^\mu+ j^\mu_\psi)
=  -  \frac{1}{\tau_q}\kappa \nabla^\mu \left( \frac{\mu}{T}\right), \quad
(u\cdot\nabla)\pi^{\mu\nu}+ \frac{1}{\tau_\pi}\pi^{\mu\nu} &= &- \frac{1}{\tau_\pi} \left(2\eta \sigma^{\mu\nu} + \zeta \Delta^{\mu\nu} \nabla\cdot u\right)\, ,
\end{eqnarray}
with $\tau_\pi$ and $\tau_q$ being additional parameters that depend on $T$ and $\mu$ (and in principle also on $\rho$). \textcolor{black}{Clearly after the timescale of the respective relaxation times, $\pi^{\mu\nu}$ and $q^\mu+ j^\mu_\psi$ will relax to their respective constitutive relations \eqref{Eq:Constitutive}.} This completes the construction of the MIS formalism for effective description of a relativistic superfluid. \textcolor{black}{Finally we note that our construction is not quite reliable at very low temperatures where higher derivative terms can contribute also to the relaxation processes. In this context, one may use the formalism discussed in \cite{Davison:2016hno}. We can choose parameters such that for generic initial conditions with supercritical temperature (and also with a range of subcritical temperatures), the system never reaches sufficiently small temperatures while undergoing Bjorken flow.}


\section{Bjorken flow}\label{Bjorken}
In order to study Bjorken flow we need to consider the Milne background
\begin{equation}
{\rm d}s^2 = - {\rm d}\tau^2 + {\rm d}x^2+ {\rm d}y^2 + \tau^2{\rm d}\xi^2
\end{equation}
and set $u^\mu = (1, \vec{0})$. Furthermore, we should consider all other variables $T$, $\mu$, $\rho$, $\psi$, $\pi^{\mu\nu}$ and $q^\mu$ to be dependent only on the proper time $\tau$. 

For a concrete example, let us consider $F$ and $V$ to be of the form given in \eqref{Eq:FV}. We further set $\mu = 0$ and $p = T^4$ as in a conformal equation of state. \textcolor{black}{Conformality implies that}
\begin{equation}\label{Eq:Cetaetc}
\eta = \frac{4}{3}C_\eta T^3, \quad \tau_\pi = \frac{C_{\tau\pi}}{T}, \quad \kappa_1 = C_{\kappa1} T,\quad \kappa_2 = C_{\kappa2} T^3
\end{equation}
where $C_\eta$, etc are constants and we set $\zeta = 0$. \textcolor{black}{Note that $C_\eta=\eta/s$.} Alternatively, we can parameterize\footnote{\textcolor{black}{In general $\eta/s$ will be a complicated function of $\sigma/T$ and $T_c/T$. The first parametrization \eqref{Eq:Cetaetc} of $\eta$ etc. is relevant at high temperatures when there is asymptotic freedom. At lower temperatures, the second parametrisation is relevant if $\sigma << T_c$.}} $\kappa_1 = C_{\kappa1} T_c$ and $\kappa_2 = C_{\kappa2} T_c^3$. None of the qualitative features that will feature in our discussion will depend on the details of such parametrizations. 
Denoting $\tau-$derivatives via a prime
, the equations of motion \eqref{Eq:EomSigma1} now take the form
\begin{eqnarray}\label{Eq:Bjorkenrho1}
\rho'' + \frac{\rho'}{\tau} + \lambda \rho^3 + \sigma(T- T_c) \rho - \rho {\psi' }^2 &=& - {C_{\kappa1}}T\rho', \\\label{Eq:Bjorkenpsi1}
( \rho^2 \tau \psi')' &=& - C_{\kappa2}T^3 \tau \psi'.
\end{eqnarray}
Furthermore, we consider a diagonal form of $\pi^\mn$:
\begin{align}
\pi^\mn =  {\rm diag}\left(0, \frac{\Pi}{2},  \frac{\Pi}{2}, -\frac{\Pi}{\tau^2}\right).\end{align}
It is convenient to use the dimensionless pressure anisotropy $\chi := 3 \Pi/4 T^4$. The conservation of the energy-momentum tensor $T^{\mu\nu}$ given in \eqref{Eq:FullTj} provides the equation for the evolution of the temperature
\begin{eqnarray}\label{Eq:TEq1}
\frac{\tau T'}{T}  =\frac{1}{3} (\chi- 1)+\sigma\frac{\rho^2 + 2 \tau \rho \rho'}{8T^3}+ \frac{\tau}{4T^3}\left(C_{\kappa1}{\rho'}^2+C_{\kappa2}T^2{\psi' }^2\right).
\end{eqnarray}
Assuming $p$ to be conformal, it is natural to modify the MIS equation \eqref{Eq:MIS} by replacing $u\cdot\nabla$ on the LHS with a Weyl-covariant derivative as in BRSSS formalism \cite{Baier:2007ix} so that the evolution of $\Pi$ \textcolor{black}{in terms of $\chi$} is given by
\begin{align}\label{Eq:ChiEq}
\tau \chi' + \frac{4}{3} \left(\chi - \frac{C_\eta}{C_{\tau\pi}} \right)+4 \chi \frac{\tau T'}{T} + \frac{\tau}{C_{\tau\pi}} \chi T = 0.
\end{align}
As such, we have a five-dimensional phase space given by $T$, $\chi$, $\psi'$, $\rho$ and $\rho'$ when we set $\mu = 0$ (for simplicity).

\section{Results}\label{results}
Superfluid Bjorken flow has formidable complexity, but the results are best understood by considering possible attractor or saddle surfaces/curves/points first. It is easy to note from our equations that we can always consistently set $\Sigma = 0$
. The evolution of $T$ and $\chi$ are then exactly same as the case of ordinary Bjorken flow discussed in \cite{Heller:2015dha}. It is well known that the system flows to a unique hydrodynamic attractor curve given by $\chi_{\rm att}(\tau T)$. 
\textcolor{black}{However, we find this curve is actually a meta-stable curve near the $\Sigma = 0$ surface: typical solutions will approach this surface before departing at late times.}

\begin{figure}
   \centering
   \includegraphics[width=0.5\textwidth]{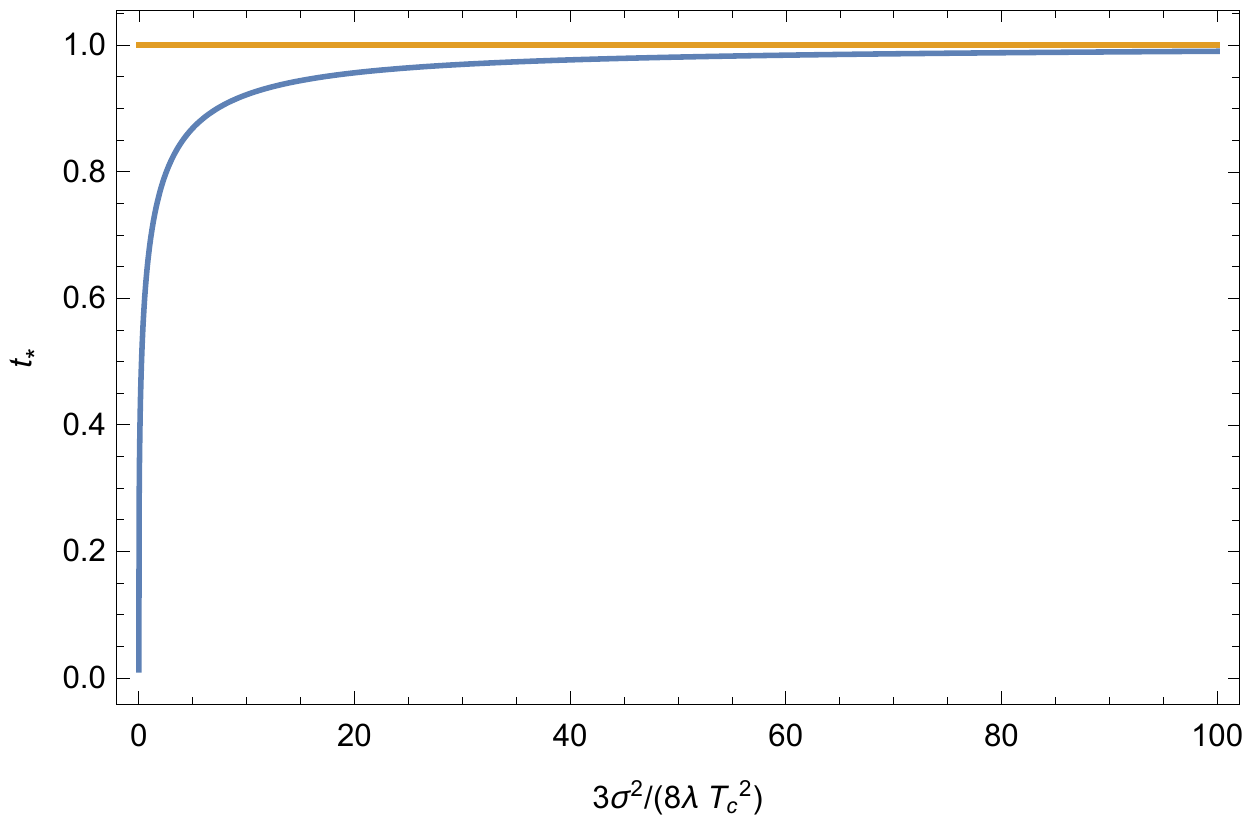}
   \caption{There is always a solution to \eqref{spec-eq} with $0\leq t_*<1$. }\label{Fig:tstar-A} 
   
\end{figure}

{In addition, we find that} the system has two fixed points! To see this we set for the moment $T =  t_* T_c$, $\chi = 0$ {(since any fixed point cannot have dissipation)} and note that the equations of motion have solutions with $\psi' =0$ and
\begin{equation}
\rho =\pm \rho_* = \pm \sqrt{\frac{\sigma T_c(1 - t_*)}{\lambda}}
\end{equation}
since $\partial V/\partial \rho = 0$ at $\rho = \pm \rho_*$. Note setting $\psi$ constant implies that we are moving on a line in the complex $\Sigma$ plane, so $\rho$ can assume any sign\footnote{\textcolor{black}{The situation is similar to motion to pure radial motion in a central potential with vanishing angular momentum. The constant radial coordinate $\phi = \phi_0$ jumps by $\pi$ once the particle crosses the origin. If we want to keep $\phi$ continuous, then we can simply allow the radial coordinate to assume both signs.}}. The equation for $T$  given by \eqref{Eq:TEq1} 
determines $t_*$ via
\begin{equation}\label{spec-eq}
 \frac{ 1 - t_*}{ t_*^3 } = \frac{8\lambda T_c^2}{3\sigma^2}\, ,
\end{equation}
in which we take $0\leq t_*< 1$. \textcolor{black}{In Fig.~\ref{Fig:tstar-A}, we see that irrespective of our choice of constants $(\sigma, \lambda, T_c)$, we are always able to find a real solution within this range.} Choosing $T_c = \sigma = \lambda = 1$, we find that $t_* \approx 0.551847$ and $\rho_* \approx   0.669442$. 

The reader may readily note that \eqref{spec-eq} orginates from a specific choice of 
the equation of state i.e. $p(T)$. Also the fixed point is determined only by the quantum effective action (the ideal component) and is independent of the relaxation terms.  The existence of such fixed points should therefore be generic in the Bjorken flow equations of the superfluid.  The reason that such fixed points with constant temperature can exist in this expanding superfluid droplet is simply because the energy gained by non-reduction of temperature is compensated by the lower energy of the expanding vacuum with broken symmetry. Generally, of course the system will have even number of such fixed points since the symmetry breaking vacuua are doubly degenerate for fixed $\psi$ and for any acceptable solution of $t_*$. At these  symmetry breaking fixed points there is no entropy production because $\rho' = \psi'  = \pi^{\mu\nu} = q^\mu = 0$. Since these are fixed points of an expanding system (with non-trivial velocity profile), these are not thermal.

 We are now ready to report our results. The system of equations for the simple case \eqref{Eq:FV} which give the evolution of ($T$, $\chi$,  $\rho$, $\psi$) under Bjorken flow with vanishing $\mu$ are  \eqref{Eq:Bjorkenrho1}, \eqref{Eq:Bjorkenpsi1}, \eqref{Eq:TEq1} and \eqref{Eq:ChiEq}. 
We recall that the phase space is five-dimensional when $\mu = 0$ with coordinates $T$, $\chi$, $\rho$, $\rho'$ and $\psi'$. The characteristics of the evolution of the system, however, depend only on whether we initialize with $T > T_c$ or $T< T_c$. 

For the purpose of generating plots, we choose $C_\eta = 1/ (4\pi)$ and  $C_{\tau\pi} = (2 - \ln 2)/(2\pi)$ as obtained by matching MIS with the holographic description of $\mathcal{N} = 4$ SYM theory at infinite 't Hooft coupling and for infinite number of colors \cite{Heller:2015dha}.  Also we choose $C_{\kappa1}= C_{\kappa2} = 1$, and $\sigma = \lambda =1$ in units $T_c = 1$. \textcolor{black}{Likewise, we will count our proper time, $\tau$, in units of $T_c=1$.} 
\begin{figure}
   \centering
   \includegraphics[width=0.48\textwidth]{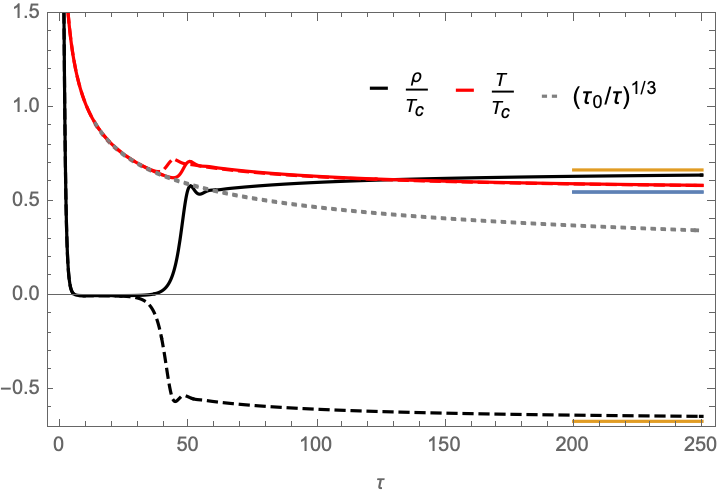}
   \hfill
 \includegraphics[width=0.47\textwidth]{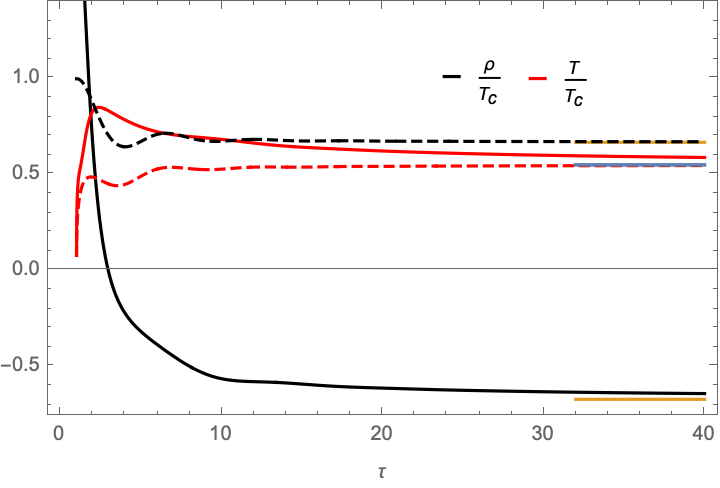}
   \caption{
   \textcolor{black}{Evolution of the condensate and temperature} for initial $T>T_c$ (left) and $T<T_c$ (right) \textcolor{black}{for two different initial values of $\rho$, distinguished by solid and dashed lines}. In the left panel with initial values of $\rho$ as 1.947 and 1.954 (corresponding to the solid and dashed lines, respectively), at early times, the temperature (red lines) quickly goes to perfect-fluid type expansion $T \approx \tau^{- 1/3}$ (shown as a gray dashed line) before switching to one of the symmetry breaking points at late time. This switching time corresponds to the rapid growth of the condensate, $\rho$, shown in black. At late times, both $T$ and $\rho$ asymptote to their respective final values, $\ts$ and $\pm\rhos$, \textcolor{black}{which we denote in both plots as blue and orange lines, respectively}. In the right panel corresponding to initial values of $\rho$ as 2 and 1 (solid and dashed lines, respectively), with initial $T<T_c$, the condensate rapidly approaches its asymptotic value. The evolution of the temperature is non-monotonic even at early times. In these plots $t_* \approx 0.551847$ and $\rho_* \approx   0.669442$ and we start with $\rho' = \psi' = 0$ and $\chi = 1$.}\label{Fig:TypicalEvolution} 
   
\end{figure}

\paragraph{Case 1 -- initially $T > T_c$:} 
In this case,
the typical evolution of the system is as shown in the left panel of Fig.~\ref{Fig:TypicalEvolution}. Early on, both $\rho$ and $\psi'$ vanish very rapidly, particularly $\rho \approx \tau^{-1/2}{e^{- \sqrt{\sigma T_c} \tau}}$. The system approaches the $\Sigma = 0$ surface  (with $\rho = \rho' = \psi' = 0$) where the evolution is purely hydrodynamic. As a result, the system gets trapped near the hydrodynamic attractor curve $\chi_{\rm att}(\tau T)$ 
where there is no symmetry breaking ($\rho$ remains exponentially small). The entrapment occurs for a long time during which the system approaches a perfect fluid type expansion ($T \approx \tau^{-1/3}$) due to its evolution close to the attractor curve. However, at late time the system suddenly reheats quickly and switches to one of the symmetry breaking fixed points with $T = t_*T_c$ and $\rho = \pm \rho_*$ (with $\psi' = 0$ and $\chi = 0$). Note that during the entire evolution of the system, the temperature remains above $t_* T_c$, including during the switching between the hydrodynamic attractor behavior and the final approach to its asymptotic value. 

\begin{figure}
   \centering
   \includegraphics[width=0.8\textwidth]{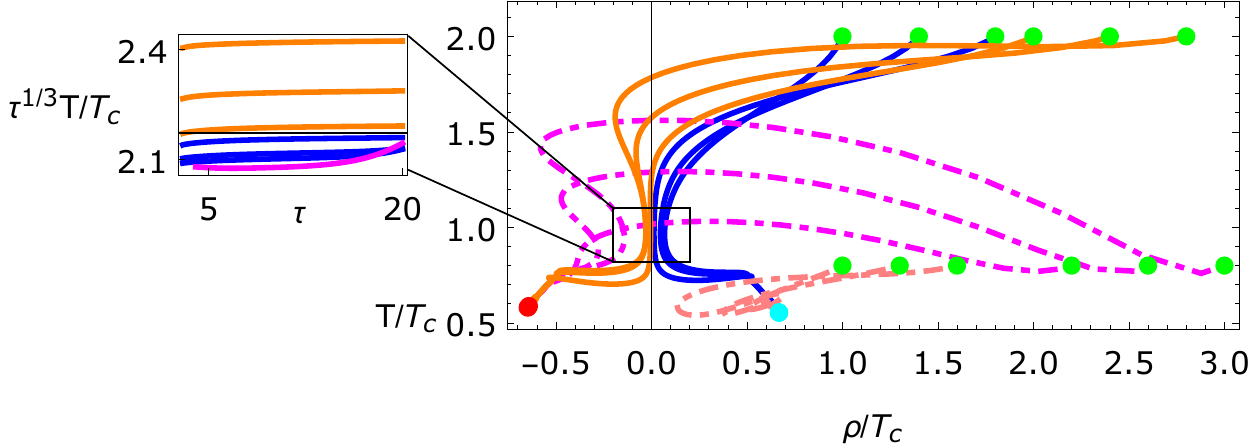}
   \caption{The evolution of the system is shown on the $T-\rho$ plane, a two-dimensional projection of the phase space. The initial values of $T$ and $\rho$ are marked by green dots and for all cases initial values of $\rho'$, $\psi'$ and $\chi$ are $0$, $0$ and $1$, respectively. At late times, the system chooses one of the two fixed points marked in red and cyan. When $T = 2 T_c$ initially, we find that the system quickly evolves to the vertical $\rho = 0$ line and gets trapped there for intermediate times. The inset plot shows that during this time $T \approx \tau^{-1/3}$ indicating that the system evolves close to the hydrodynamic attractor curve on $\Sigma = 0$ surface (with $\rho =\rho' = \psi' = 0$). If we  start with $T = 0.8 T_c$, then the system does not get trapped by the hydrodynamic attractor curve but merely passes through the vertical $\rho = 0$ line (except if we start close to the border separating the basin of attraction of the two fixed points, as shown by the magenta line in the inset). } \label{Fig:PhaseSpace}
   \label{}
\end{figure}

In Fig.~\ref{Fig:PhaseSpace}, we further illustrate the above by fixing 
the initial temperature to $2T_c$ and initializing with $(\rho',\psi',\chi)=(0,0,1)$, but varying the initial value of $\rho$. 
As mentioned above, the qualitative features do not change if we choose initial values of $T$, $\rho'$, $\psi'$ and $\chi$ differently as long as $T > T_c$ initially. For any initial value of $\rho$, the system gets trapped close to the vertical $\rho = 0$ line, representing the hydrodynamic attractor curve where we have perfect-fluid like expansion $T \approx \tau^{-1/3}$ for a considerable time (see the inset) until it switches to one of the symmetry breaking fixed points marked with red and cyan colors. Varying initial values of $\rho$, we find alternating intervals in which the system chooses one of these fixed points at late times. The switching time depends on the initial condition and increases with the inital value of $T$ very rapidly. 

\begin{figure}
   \centering
   \includegraphics[width=0.4\textwidth]{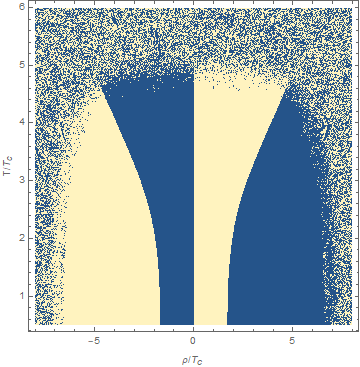}
   \caption{The projection of the basin of attractions of the two fixed points on the $T-\rho$ plane are shown here. The color coding is yellow for $+\rhos$ and blue for $-\rhos$. Note that below a certain value of temperature, there is a clear separation between a choice of initial conditions determining which $\rhos$ the system will evolve to. This is not the case for higher temperatures and higher $\vert \rho \vert$, where predictability from the initial conditions becomes difficult. The boundary of the basins of attraction is possibly a  fractal at higher tempertures.}\label{Fig:Basin}
   \label{}
\end{figure}

To study the basin of attraction of the fixed points, we fix the initial values 
$(\rho',\psi',\chi)=(0,0,1)$ and vary initial values of $T$ and $\rho$.  We readily observe from Fig. \ref{Fig:Basin} that for initial values of $T $ and $\vert \rho\vert$  greater than about $5 T_c$ and $ 7 T_c$, respectively, the basin of attraction of these two fixed points (marked with yellow and blue for $\rho_*$ and $-\rho_*$, respectively) are interlaced in a complex manner, making long term prediction difficult. It is likely that the basin boundaries are fractal in these regimes. 

We emphasize that even if the system is initialized with non-vanishing $\rho'$ and $\psi'$, it rapidly approaches the hydrodynamic attractor curve with vanishing $\Sigma$ and lingers there for a long duration of time before rapidly switching to one of the fixed points. The basin of attraction of the fixed points has similar features.

If we start from the boundary of the basin of attraction of the two fixed points initially, then the system should get trapped by the hydrodynamic attractor curve on the vanishing $\Sigma$ surface (with $\rho = \rho' = \psi' = 0$) and would asymptotically approach it without switching to any of the fixed points. 

\paragraph{Case 2 -- starting with $T \leq T_c$:} The evolution is qualitatively different to the previous case, although the system settles to one of the fixed points for a generic initial condition. As shown in Fig. \ref{Fig:TypicalEvolution}, the temperature behaves non-monotonically at initial time even if we start with $T \approx t_*T_c$. The condensate $\rho$ approaches $\pm \rho_*$ while $\psi'$ and $\chi $ vanishes over the same timescale. Thus, the system approaches one of the fixed points without getting trapped near $\Sigma = 0$, unless the initial condition is close to the boundary of the basin of attraction as demonstrated in Fig. \ref{Fig:PhaseSpace}. In the latter case, the system evolves close to the hydrodynamic attractor for a certain period of time. This is expected because if we start from the boundary of the basins of attraction, then the system should approach the hydrodynamic attractor curve in the long run instead of evolving to one of the fixed points. We readily observe from Fig. \ref{Fig:Basin} that the projections of the basin of attraction on $T-\rho$ plane separates into two simply connected regions.

\textcolor{black}{In the following section, we analyze the MIS theory about thermal equilibrium and argue that the fixed points are actually unstable against inhomogeneous perturbations. The fate of this instability can be studied using our MIS framework.}


\section{Linearized fluctuations}\label{lin-pert}

Here, we compute the linearized fluctuations around thermal equilibrium at zero chemical potential. Note that in this section we are not considering an expanding fluid and we use the standard Minkowski coordinates. We show that the broken phases with $T \leq t_* T_c < T_c$ (with $t_* T_c$ corresponding to the value of the temperature at the fixed point of the Bjorken flow) are unstable in our MIS theory. While the non-hydrodynamic relaxation poles remain on the lower half-plane of complexified frequency, the speed of sound, the diffusion constant and the sound attenuation coefficient change sign at $T = t_*T_c $ and thus are negative for $T < t_* T_c$. This hydrodynamic instability should lead to spinodal decomposition \cite{Boyanovsky:2006bf,PhysRevResearch.2.033138}.

Without loss of generality, we take the momentum of the fluctuation in the $z$-direction. \textcolor{black}{After recasting the equations of motion, \eqref{Eq:Bjorkenrho1} and \eqref{Eq:Bjorkenpsi1}, energy momentum tensor conservation \eqref{Eq:TEq1} and MIS equation \eqref{Eq:ChiEq} in Minkowski coordinates and linearizing}, we see that these equations are linear in $V\equiv(\delta T, \delta \rho, \delta \psi, \delta u, \delta \pi)$ in Fourier space. We can recast the equations in matrix form $Q(\omega,k)\cdot V=0,$ where $Q$ is a $5\times 5$ matrix. We find the dispersion relations by setting the determinant of $Q$ to zero.
The perturbations of the hydrodynamic variables are given by
\begin{align}
\label{shear-pert}
&\text{shear:} \quad \pi^{zx}  \sim \delta \pi^{zx} e^{-i \omega t+i kz} \quad \text{and} \quad  u^x  \sim  \delta u^x e^{-i \omega t+ i kz},\\
\label{sound-pert}
&\text{sound:} \quad T \sim  T_0 + \delta T e^{-i \omega t+ i kz},\quad\pi^{zz}  \sim  \delta \pi^{zz} e^{-i \omega t+ i kz} \quad \text{and} \quad 
u^z  \sim  \delta u^z e^{-i \omega t+ i kz}.
\end{align} 
The fluctuations in the scalar sector are
\begin{align}
\rho\sim  \rho_0+\delta  \rho e^{-i \omega t+ i kz} \quad \text{and} \quad \psi  \sim  \psi_0 + \delta \psi  e^{-i \omega t+ i kz}.
\end{align}
We distinguish between the broken and unbroken phase by the equilibrium value of the condensate. 
\subsection{\textcolor{black}{Unbroken phase}}
When $\rho_0=0,$ we are in the unbroken phase, $T>T_c$. In this case, the determinant of $Q$ factors neatly into the phase, condensate and hydrodynamic dispersions. Since $\rho_0$ vanishes, the dispersion of the phase is trivial. We find the expected sound and shear modes of a conformal system (with $\eta = C_\eta T_0^3$)
\begin{align}\label{unbroken-sound}
\omega_{sound} &= \pm  \frac{ k }{\sqrt{3}} -  \frac{ i \eta  k^2}{2 T_0^4}+\mathcal{O}\left(k^3\right), \quad \text{and} \quad
      \omega_{shear}= -\frac{3i\eta  k^2}{2T_0^4}+\mathcal{O}(k^3),
   \end{align}  
   as well as the non-hydrodynamic modes: one from the MIS equation and two from the scalar field
\begin{align}\label{MIS-mode}
\omega_{relaxation} &=  -\frac{i}{\tau_\pi}+\mathcal{O}(k^2), \\
\omega_{scalar} &=  - \frac{1}{2} \left(i \kappa_1 \pm \sqrt{4 \sigma  ( T_0-T_c)-\kappa_1^2}\right) + \mathcal{O} \left(k^2 \right),
   \end{align}
   respectively. Note that the superfluid mode is overdamped when $\kappa_1 > \sqrt{2\sigma  ( T_0-T_c)}$.

\subsection{\textcolor{black}{Broken phase}}
Next, we turn our attention to the broken phase, $ T<T_c  $ , and $ \rho_{0} =\pm \sqrt{\frac{\sigma (T_c- T_0)}{\lambda}}$. 
The dispersion relation of the phase is now no longer trivial, indicating the presence of a Goldstone mode
\begin{align}
\rho_0^2(\omega^2-k^2) +i \kappa_2 \omega=0.
\end{align}
Clearly we get two modes which at low $k$ take the form
\begin{equation}
\omega_{diffusion} = -i \frac{\rho_0^2}{\kappa_2} k^2 + \mathcal{O}(k^4), \quad \omega_{relaxation} = - i \frac{\kappa_2}{\rho_0^2} +\mathcal{O}(k^2).
\end{equation}
Thus one behaves as a diffusion mode and the other as a relaxation mode in both sound and shear channels.

The shear sector in the broken phase with perturbations \eqref{shear-pert} has the following determinant 
\begin{align}
0=
\left(w^2- k^2-
m_\sigma^2 
   +i \kappa_1 \omega \right)
   \left(
   T_0 \omega (\tau_\pi \omega+i) \left(8 T_0^3-3 \rho_0^2 \sigma \right)-6\eta  k^2\right) ,
\end{align}
where we introduced the mass of the condensate, $m_\sigma^2:=3
   \lambda  \rho_0^2+\sigma  (T_c-  T_0)=4 \sigma (T_c-T_0).$
The determinant factorizes into the condensate and hydrodynamic modes. For small $k$, we see that the shear modes are 
\begin{align}
\omega_{QNM}&=\frac{1}{2}\left(-i\kappa_1\pm \sqrt{4m_\sigma^2-\kappa_1^2}\right)+\mathcal{O}(k^2),\\
\omega_{shear} &= -\frac{6 i\lambda  \eta  k^2}{8 \lambda T_0^4-3  \sigma^2(T_c-T_0) T_0}+\mathcal{O}(k^3),
\end{align}
as well as the MIS mode \eqref{MIS-mode}. \textcolor{black}{We call the first mode quasi-normal mode (QNM) simply because it has both real and imaginary parts at zero momentum (no analogy with black hole QNMs are implied here).} We note that the diffusion constant has a pole in $T_0$ precisely when $T_0=t_* T_c$ and its sign changes as we further lower $T_0$. A negative diffusion constant leads to what is known as uphill diffusion (against the concentration gradient) enhancing inhomogeneities and thus spinodal decomposition. As a side remark, the two scalar (condensate) modes become one diffusive mode in the limit $m_\sigma^2\rightarrow 0$ (see the Goldstone sector discussed above), which in the context of the chiral phase transition can be thought of as the transition from pion propagating modes to the diffusive quark mode \cite{derek_tbp}. Of course, in the present context, we are working with a complex scalar field and not attempting to capture the $O(4)$ dynamics relevant for the chiral phase transition.

\begin{figure}
   \centering
   \includegraphics[width=0.48\textwidth]{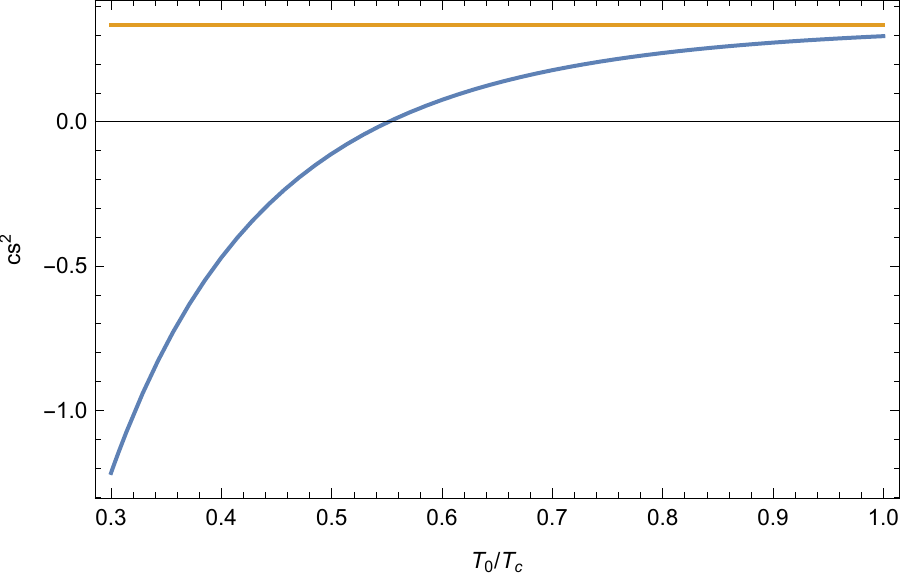}
   \hfill
 \includegraphics[width=0.47\textwidth]{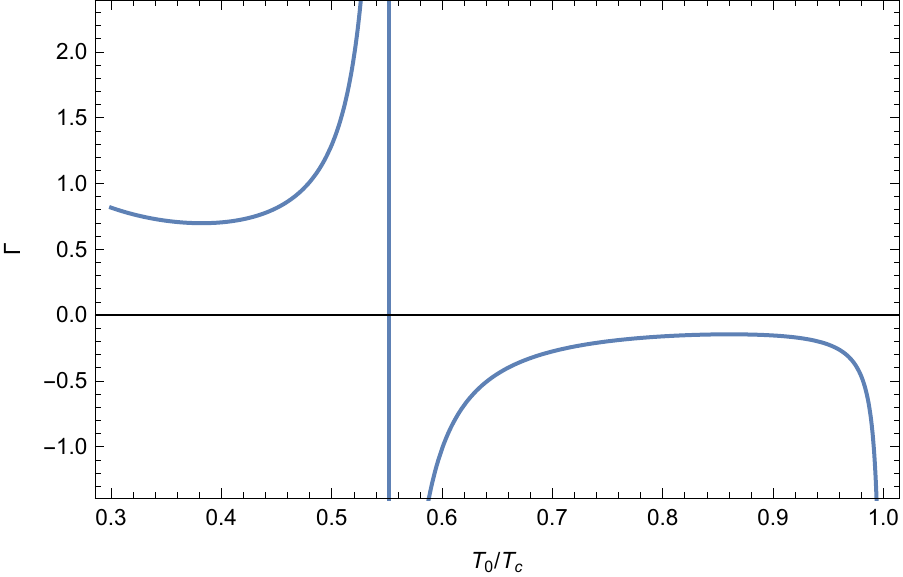}
   \caption{Left: the square of the speed of sound as a function of equilibrium temperature. As $T_0\rightarrow t_*$, the speed of sound vanishes, while as $T_0$ approaches the critical temperature $T_c$, it grows and monotonically reaches the conformal value, $1/3.$ Right: the sound attenuation coefficient as a function of equilibrium temperature, which grows negative and large as $T_0$ approaches $t_* T_c$. }\label{Fig:broken-sound} 
   
\end{figure}

Finally, in the sound channel with perturbations \eqref{sound-pert}, the dispersion takes the following form
\begin{align}
0&=48 \eta  k^2 \lambda  T_0 \omega \left(4 k^2 \lambda  T_0^2-8 \lambda  \sigma  T_0^3-4 \lambda 
   T_0^2 \left(-2 \sigma  T_c+w^2+i \kappa_1 w\right)-\sigma ^3 T_0+\sigma ^3
   T_c\right)\nonumber\\
   &-(\tau_\pi w+i) \left(8 \lambda  T_0^3+3 \sigma ^2 T_0-3 \sigma ^2
   T_c\right) \Big{[}-k \left(8 k \lambda  T_0^4+3 k \sigma ^2 T_0 (T_0-T_c)\right)
   \left(k^2+2 \sigma  (T_c-T_0)-w^2-i \kappa_1 w\right)\nonumber\\
   &-6 T_0^2 w^2 \left(-4 k^2
   \lambda  T_0^2+8 \lambda  \sigma  T_0^3+4 \lambda  T_0^2 \left(-2 \sigma  T_c+w^2+i
   \kappa_1 w\right)+\sigma ^3 T_0-\sigma ^3 T_c\right)\Big{]}
\end{align}
We see that in the small $k$ limit, we have the following modes
\begin{align}  \label{broken-sound}
   \omega_{sound} &=\pm c_s k-i \Gamma k^2 +\mathcal{O}(k^{3}),\quad c_s^2 := \frac{8 \lambda  T_0^3-3 \sigma ^2 (T_c- T_0)}{24 \lambda  T_0^3+3
   \sigma ^2 T_0},  \nonumber\\ 
   \Gamma &:=\frac{4 \eta  \lambda }{8 \lambda  T_0^4-3 \sigma ^2 (T_c-T_0)T_0}
   +\frac{ \kappa_1 \sigma  \left(8 \lambda  T_0^3-3 \sigma ^2 (T_c-
   T_0)\right)}{12 T_0 (T_c-T_0) \left(\sigma ^2+8 \lambda  T_0^2\right)^2} ,\\
\omega_{QNM} &=\frac{1}{2}\left(-i \kappa_1\pm
\sqrt{-\kappa_1^2 +8 \sigma   (T_c-T_0)+\frac{\sigma ^3
   (T_c-T_0)}{\lambda T_0^2 }}
   \right)
   +\mathcal{O}(k^2).
\end{align}
Above, the first mode is clearly the sound mode but the square of the speed of sound $c_s^2$ becomes negative for $T_0 < t_*T_c$ triggering spinodal decomposition. In Fig.~\ref{Fig:broken-sound}, we plot $c_s^2$ and the sound attenuation coefficient $\Gamma$ which behaves similarly as the diffusion coefficient in the shear sector, diverging at $T_0 = t_* T_c$ and becoming negative as we further lower $T_0$.

We note that the instabilities occur only in the hydrodynamic sound and shear modes only for $T_0 \leq t_*T_c$. This indicates that the fixed points of the Bjorken flow could be unstable against inhomogeneous perturbations.
 However, 
 when we start with supercritical temperatures (and a range of subcritical temperatures), the temperature remains above $t_* T_c$ for its entire evolution for a large class of initial conditions. We would like to investigate the fate of the instabilities of the fixed points using an appropriately modified MIS framework in the future.

\section{Discussion}\label{discussion}
In this work, we setup the MIS formalism for superfluids, extending the quantum effective action approach and study the Bjorken flow. We find that the dynamics of the system is governed by the existence of the conventional hydrodynamic attractor with unbroken symmetries and new fixed points determined by the potential for the condensate and the equation of state of the system. Particularly, if the initial temperature is above $T_c$, we find that the Josephson condition is satisfied by the phase very rapidly while the condensate becomes exponentially small over a similar timescale and undergoes slow-roll during which the system gets trapped in the vicinity of the conventional hydrodynamic attractor. This persists for a long time until the system reheats and switches over to one of the symmetry-breaking fixed points. If the initial temperature is below $T_c$, the system does not generically get trapped by the conventional hydrodynamic attractor, but rather evolves to one of the fixed points. In both cases, the hydrodynamic attractor traps the system forever if we start with initial conditions at the border of the basin of attraction of the fixed points.  

As discussed before, we expect that our key result that the superfluid Bjorken flow is governed by a combination of hydrodynamic attractor and an even number of non-dissipative fixed points should be a generic phenomenological feature. The fixed points are determined by the ideal component of the dynamics obtained from the quantum effective action with the relaxation mechanism playing no role while the hydrodynamic attractor appears in any strong coupling or weak coupling description of thermalization. 

\textcolor{black}{The analysis of fluctuations about thermal equilibrium indicate that the system has unstable hydrodynamic modes when the temperature corresponds to values lower than that at the fixed points.  This indicates that the fixed points are unstable against spinodal decomposition. In the future, we would like to investigate this phenomenon by introducing inhomogeneities in the initial conditions  and using our MIS theory with necessary improvements.}


It would be important to study the Bjorken expansion of the superfluid in a UV-complete setup at strong coupling, particularly in a holographic model. 
Of particular interest would be to understand if the approach to the hydrodynamic attractor and subsequent switching to the symmetry breaking fixed points occurs rapidly as in the MIS setup.  

\textcolor{black}{Furthermore, to be of relevance to the quark-gluon plasma we will need to consider interacting multi-component systems as  in \cite{Schmitt:2014eka}. Our model is essentially that of the quark sector. The gluonic degrees of freedom should be introduced as a separate fluid. Furthermore, it may be necessary to include  both weakly coupled and strongly coupled degrees of freedom. Such hybrid fluid models with a higher dimensional hydrodynamic attractor has been discussed in \cite{Kurkela:2018dku,Mitra:2020mei}. In case of a multicomponent system, the reheating transition of the superfluid to the fixed point may not be possible or may not occur irreversibly because of the transfer of energy to the other components of the full system.}


Also, it would be interesting to explore more general forms of expanding flows, particularly those with spherical symmetry (see \cite{Endlich:2012pz} for a relevant model). As the superfluid condensate is naturally driven to slow roll on a hydrodynamic attractor with a natural mechanism of exit from the slow-roll, while reheating of the system upon switching to one of the fixed points, this raises the possibility that one can use 
such a superfluid as an inflaton in cosmology with a natural preheating mechanism in place.

\begin{acknowledgments}
 \textcolor{black}{It is a pleasure to thank Matteo Baggioli, Eduardo Grossi, Alexander Haber, Arul Laksminarayan, David M{\"u}ller, Anton Rebhan and Derek Teaney  for helpful discussions. We also thank Anton Rebhan for comments on the manuscipt.} AM acknowledges support from the Ramanujan Fellowship, ECR award of the Department of Science
and Technology of India and the New Faculty Seed
Grant of IIT Madras. AS is supported by the Erwin Schr{\"o}dinger Fellowship from Austrian Science Fund (FWF), project no. J4406.
\end{acknowledgments}

\bibliography{ssb-refs}

\end{document}